# Magnetic Force Microscopy for Nanoparticle Characterization

Gustavo Cordova[1], Brenda Yasie Lee[1] and Zoya Leonenko[1,2,3*]

[1]*Department of Biology, University of Waterloo, Waterloo ON, Canada*
[2]*Department of Physics and Astronomy, University of Waterloo, Waterloo ON, Canada*
[3]*Waterloo Institute for Nanotechnology, University of Waterloo, Waterloo ON, Canada*

[*]**Correspondence to:**
Dr. Zoya Leonenko
Department of Physics and Astronomy,
University of Waterloo,
200 University Avenue West, Waterloo,
Ontario, Canada
Tel: 519-888-4567 x 38273
E-mail: zleonenk@uwaterloo.ca

## Abstract

Since the invention of the atomic force microscope (AFM) in 1986, there has been a drive to apply this scanning probe technique or a form of this technique to various disciplines in nanoscale science. Magnetic force microscopy (MFM) is a member of a growing family of scanning probe methods and has been widely used for the study of magnetic materials. In MFM a magnetic probe is used to raster-scan the surface of the sample, of which its magnetic field interacts with the magnetic tip to offer insight into its magnetic properties. This review will focus on the use of MFM in relation to nanoparticle characterization, including superparamagnetic iron oxide nanoparticles, covering MFM imaging in air and in liquid environments.

## Abbreviations

AFM: Atomic Force Microscopy; HL-60: Human Promyelocytic Leukemia Cells; MCF-7: Michigan Cancer Foundation-7 (breast cancer cell line); MFM: Magnetic Force Microscopy; PMMA: Poly(methyl methacrylate); PS: Pulmonary Surfactant; SMNP: Supramolecular Magnetic Nanoparticles; SPION: Superparamagnetic Iron Oxide Nanoparticles; SQUID: Superconducting Quantum Interference Device

## Introduction

Magnetic force microscopy (MFM) is a type of scanning probe microscopy where a magnetic probe is used to scan the sample surface while interacting with the magnetic fields of the sample [1, 2]. The most common method is called the "Two- Pass Technique" or "Hover Mode Scanning Method" [3-5], in which the sample is scanned twice (see Figure 1): once to produce a topographical image (AFM) and a second time to produce an MFM image. This is the most commonly used method for MFM, because it is relatively simple to implement and will provide high phase contrast in ambient condition [6].

On the first trace of the sample, the cantilever directly scans the surface of the sample just as it would for AFM in its intermittent contact or "tapping" mode, producing a topographical image. During this time, short range interactions (e.g. Van der Waals forces) have the most significant effect on the cantilever. Subsequently on the retrace, the cantilever is raised to a user-defined height (eg. 50 nm) away from the surface and scans for the magnetic signal by following the topographical pattern from the previous trace as seen in Figure 1A. Long range interactions, such as magnetic forces, are the most prevalent in the retrace and thus allow the MFM image to reflect the magnetic properties of the sample being imaged.

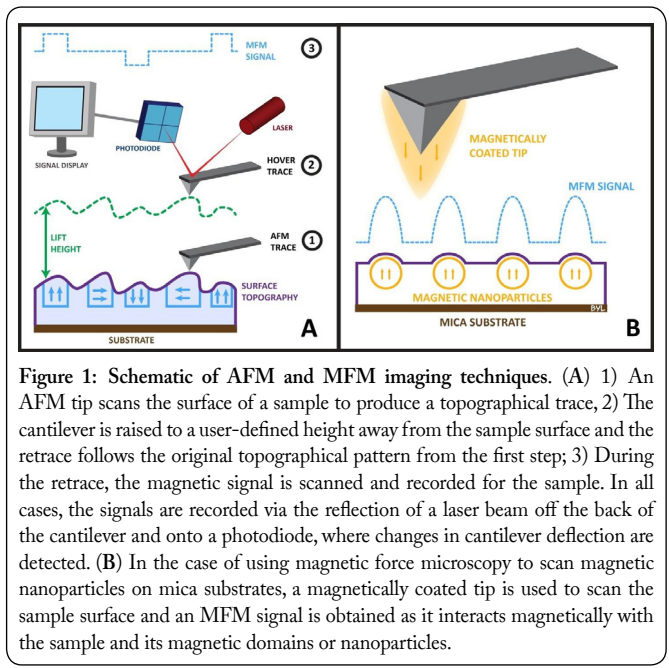

**Figure 1: Schematic of AFM and MFM imaging techniques.** (**A**) 1) An AFM tip scans the surface of a sample to produce a topographical trace, 2) The cantilever is raised to a user-defined height away from the sample surface and the retrace follows the original topographical pattern from the first step; 3) During the retrace, the magnetic signal is scanned and recorded for the sample. In all cases, the signals are recorded via the reflection of a laser beam off the back of the cantilever and onto a photodiode, where changes in cantilever deflection are detected. (**B**) In the case of using magnetic force microscopy to scan magnetic nanoparticles on mica substrates, a magnetically coated tip is used to scan the sample surface and an MFM signal is obtained as it interacts magnetically with the sample and its magnetic domains or nanoparticles.

## Theory and Advances

Magnetic force microscopy was developed in 1987 as a new method for imaging magnetic fields down to a resolution of 100 nm [7]. This novel method used the magnetic force gradient acting between a magnetic tip and magnetic sample surface to map the magnetic field distributions on the nanoscale. The instrumentation setup involved a 25 μm diameter iron wire which served as both the cantilever and the tip, where one side of the wire was tapered down to a diameter of 2 μm with a 0.1 nm diameter tip etched at its end. The resolution achieved was 100 nm. Although the operating principle of MFM is identical to AFM, the interactions between the tip and the sample are of magnetic nature and are different from forces acting in AFM.

MFM measurements are commonly taken in the Two-Pass Technique, where AFM topography is measured in the first pass, and the second pass involves the retrace of the first pass at a small *z*-offset to record long-range magnetic forces. As the tip oscillates, a decaying magnetic force is experienced as the tip-sample separation increases; it is this force gradient that changes the phase of the cantilever's oscillation and allows for a phase signal to be recorded during the MFM measurement. The phase signal ($\xi$) can be calculated using the following formula, which is derived from the application of the Euler beam theory to the force gradient ($f_m$) and phase shift ($Q \cdot f'_m \cdot k^{-1}$) [8]:

$$\delta\xi = \gamma_1 \frac{Q}{k} f'_m$$

Where $\gamma_1$ = 0.97, a constant that originates from the analysis of the thermal motion of a cantilever [9].

Due to its obvious application to magnetic materials research, MFM became a popular method in this field. However, there were several shortcomings during its early years, specifically with regards to its resolution and sensitivity to achieve measurements below 100 nm [10]. In order to improve the spatial resolution and signal strength of the MFM technique, different methods and advancements were tested to optimize the magnetized portion of the probe, the tip-sample distance, and the sample surface. Techniques emerged to fabricate advanced MFM tips or probes with tip sizes as small as 5 nm in diameter using electron beam deposition, focused ion beam milling, and attaching fullerene carbon multi walled nanotubes. Minimizing the size of the force-sensitive component of the probe (the tip apex) and keeping the tip close to the sample surface would help to improve lateral resolution [10]. Even with increased resolution, the interpretation of MFM signals and images is complex and requires further understanding of the types of interactions occurring between the tip and the sample. A recent study was performed to determine which sources of interaction contributed the most to the MFM signal, and it was discovered that the observed MFM response was mainly from electrostatic interactions and not with the magnetism of the sample itself [11]. As a result, correlating the MFM signal with the magnetism of the sample will require further extensive analysis of the different interactions that play a role between the tip and sample.

The bimodal AFM technique has been used to corroborate the theory proposed by Rodriguez and Garcia that adding a second resonant mode with a bimodally excited cantilever would be highly sensitive to long-range forces such as van der Waals interactions [12, 13]. This bimodal MFM method allows for one-pass topographical and magnetic force imaging with high spatial resolution, with image quality equivalent to that of the standard two-pass MFM. This technique has also been combined with the use of an external magnetic field to excite the magnetic tip of the probe to detect various magnetic materials. It was demonstrated that its sensitivity and signal-to-noise ratio were superior to conventional MFM techniques [14]. In both studies, it was observed that this technique also had increased spatial resolution when the tip was closer to the sample surface [12, 14].

## Advantages and Applications

MFM offers certain advantages over other imaging techniques, which has led to its increasing prevalence as a tool for magnetic nanoparticle characterization [15-20]. It has the capability to separate the magnetic interactions from the other tip sample forces (such as van der Waals, and other forces recorded in AFM). MFM can be used to detect nanoscale magnetic domains as well as simultaneously obtain both AFM phase and topography images, something that is not possible using more traditional nanoscale imaging techniques such as transmission or scanning electron microscopy. In our recent study [3], we examined the application of MFM for the detection and localization of superparamagnetic iron oxide nanoparticles (SPIONs). Figure 2 shows the corresponding MFM phase and topographical images for bare as well as silica-coated SPIONs. A study in the Universitat de Barcelona in Spain demonstrated that MFM could be used to distinguish between superparamagnetic and blocked states in aggregates of iron oxide nanoparticles [21]. Another study used MFM to investigate the magnetic microstructure of nanoparticles with single- and multi-domain octahedral particles [22].

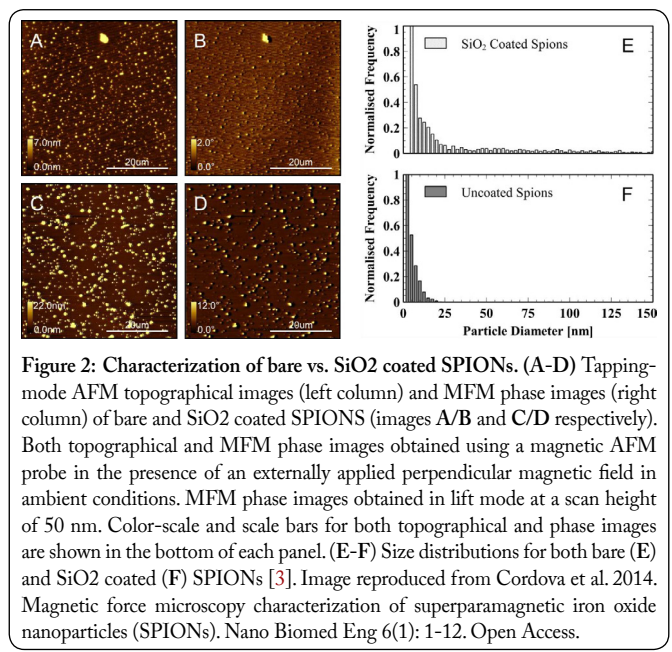

**Figure 2: Characterization of bare vs. SiO2 coated SPIONs. (A-D)** Tapping-mode AFM topographical images (left column) and MFM phase images (right column) of bare and SiO2 coated SPIONS (images **A/B** and **C/D** respectively). Both topographical and MFM phase images obtained using a magnetic AFM probe in the presence of an externally applied perpendicular magnetic field in ambient conditions. MFM phase images obtained in lift mode at a scan height of 50 nm. Color-scale and scale bars for both topographical and phase images are shown in the bottom of each panel. (**E-F**) Size distributions for both bare (**E**) and SiO2 coated (**F**) SPIONs [3]. Image reproduced from Cordova et al. 2014. Magnetic force microscopy characterization of superparamagnetic iron oxide nanoparticles (SPIONs). Nano Biomed Eng 6(1): 1-12. Open Access.

Perhaps most notably, MFM is capable of operating under ambient conditions, at varying temperatures, ultra-high vacuums, as well as liquid environments [3, 23] while still providing resolution to less than ten nanometers. This is essential because it allows magnetic nanoparticles to be localized and characterized *in vitro* [24] or inside polymer films and cell-like systems [3, 25, 26]. Because of the pervasiveness of magnetic nanoparticles in biomedical applications, it is crucial that when these nanoparticles are characterized, it is done so under conditions relevant to their application (e.g. a physiologically relevant environment, most importantly inside the cells). Until recently, the potential of MFM in this regard has been largely unexplored. Figures 3 and 4 shows separate studies where various cell lines were incubated with magnetic nanoparticles, both imaged using MFM in ambient conditions.

Liquid imaging of SPION nanoparticles was also reported by Cordova et al. in 2014 [3] using the Two-Pass Technique (Figure 5). Until this point, the majority of MFM imaging of magnetic nanoparticles had only been undertaken in ambient conditions. The single-pass bimodal MFM technique had also been successfully used to detect and image superparamagnetic nanoparticles in liquid down to 5 nm [27]. Imaging magnetic nanoparticles in liquid served as a proof of principle for using MFM to image magnetic nanoparticles within cells in order to better understand their distribution and behavior in a physiologically relevant environment. Another recent study published from the Universidad Autonoma de Madrid in Spain reported successful MFM imaging of nanoparticles in liquid [23]. In this study, MFM was used to not only image these nanoparticles, but to optimize the MFM signal acquisition in liquid media so that it could be applied to the characterization of magnetic nanoparticles and their magnetic signals. They were able to image the magnetic domains of hard disk drives as well as DMSA-coated $Fe_3O_4$ ferromagnetic nanoparticles in both air and liquid environments (Figure 6) [23]. To maximize the MFM signal in drive amplitude modulation mode for AFM, the group varied relevant measuring and operating conditions such as cantilever oscillation amplitude, set point, and Z lift distance. The authors [23], observed that the magnetic signal decreased with tip-sample distance, as opposed to previous study [3], where slight increase was observed within the distances studied. This discrepancy may originate due to coupling between the electrical and magnetic signal. More research will need to be performed in relation to MFM imaging in liquid and on effect of tip-sample distance on the cross-coupling between the magnetic and electrical signals.

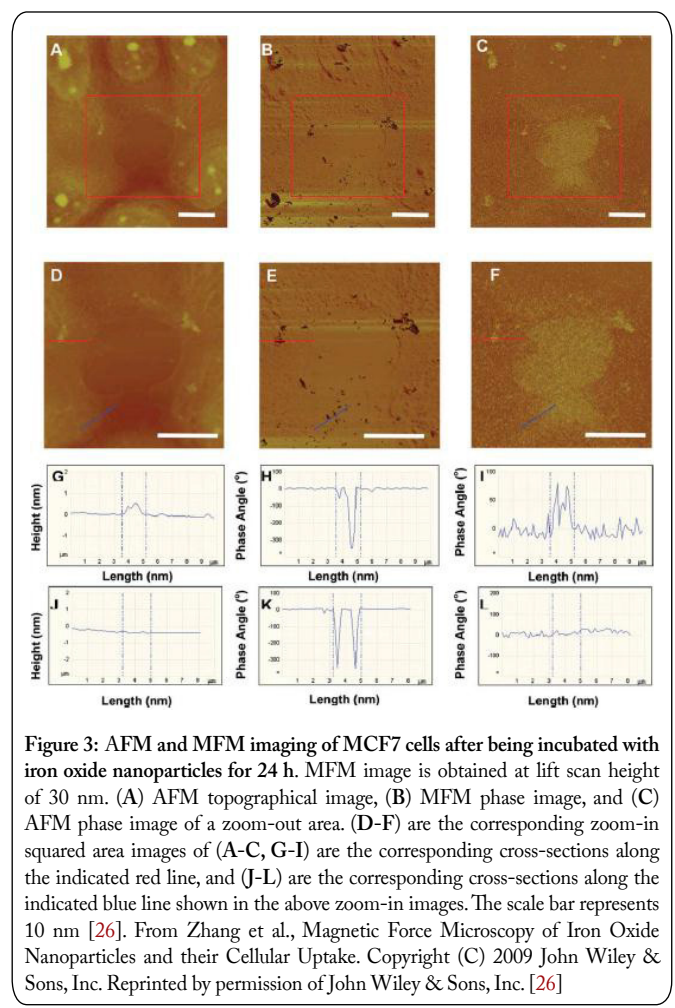

**Figure 3: AFM and MFM imaging of MCF7 cells after being incubated with iron oxide nanoparticles for 24 h**. MFM image is obtained at lift scan height of 30 nm. (**A**) AFM topographical image, (**B**) MFM phase image, and (**C**) AFM phase image of a zoom-out area. (**D-F**) are the corresponding zoom-in squared area images of (**A-C**, **G-I**) are the corresponding cross-sections along the indicated red line, and (**J-L**) are the corresponding cross-sections along the indicated blue line shown in the above zoom-in images. The scale bar represents 10 nm [26]. From Zhang et al., Magnetic Force Microscopy of Iron Oxide Nanoparticles and their Cellular Uptake. Copyright (C) 2009 John Wiley & Sons, Inc. Reprinted by permission of John Wiley & Sons, Inc. [26]

In terms of biological applications, MFM has also been used in studying the magnetic properties of biomolecules, magnetotactic bacterium [28] and biogenic nanoparticles [29]. Magnetic nanoparticles naturally occurring in biological systems have also been studied, including iron compounds associated with neurological disorders [30], magnetic domains in magnetotactic bacteria [31] and iron deposits in livers with Hepatitis B [32].

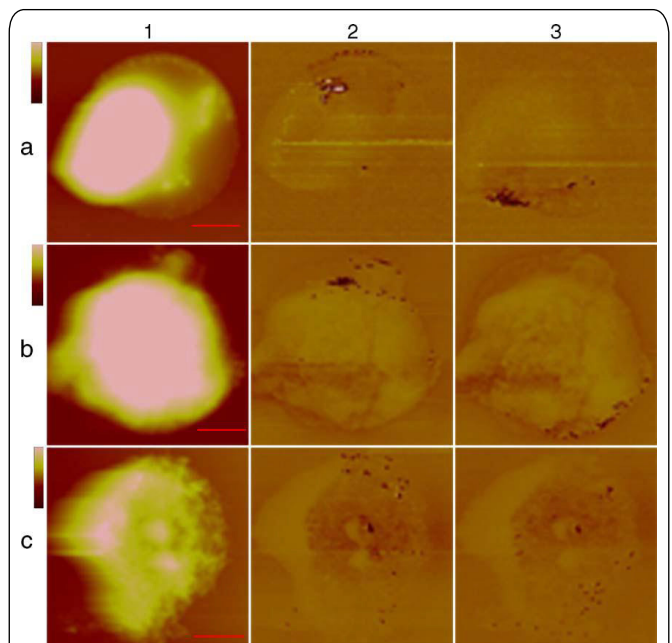

**Figure 4: MFM images of HL-60 cell treated with SOs-coupled SMNPs after 48 h and 72 h respectively**. Column 1 (far left) is a morphological image while Columns 2 and 3 are magnetic images. Column 2 shows the tip scanned from the tip to the bottom while Column 3 shows the tip scanning from the bottom to the top. Scale bars are (**a**) 3.5, (**b**) 2.4 and (**c**) 4.3 μm. Color bars are (**a**) 1.5, (**b**) 2.5 and (**c**) 1.5 μm. From the morphological image, the cell displayed different extents of apoptotic characteristic morphology. From the magnetic image, the particles can be clearly seen in the cell [25]. Reprinted from Biophysical Chemistry, Volume 122, Issue 1, Shen H et al., Magnetic Force Microscopy Analysis of Apoptosis of HL-60 Cells induced by Complex of Antisense Oligonucleotides and Magnetic Nanoparticles, pp 1-4, Copyright (C) 2006, with permission from Elsevier. [25]

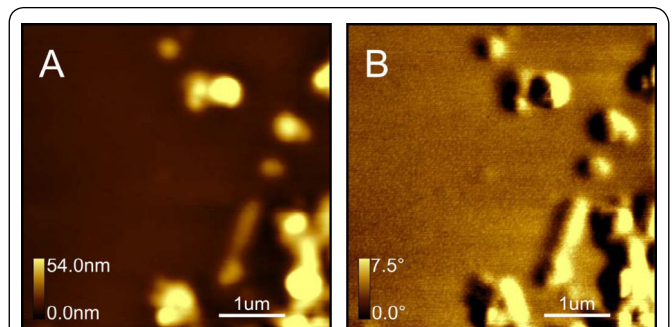

**Figure 5: MFM image (in liquid) of SiO2 coated SPIONs spin-coated with PMMA**. Tapping-mode AFM topographical image (A) and MFM phase image (B) of SiO2 coated SPIONs spin-coated with PMMA taken at a lift height of 50 nm. Images were taken using a magnetic AFM probe in the presence of an externally applied perpendicular magnetic field in liquid (water). A spin motor with an applied voltage of 1V for 15 seconds was used to spin coat a 3% solution of PMMA (in toluene) in order to cover the SPION sample with ~30 nm of PS. Color-scale and scale bars are shown in the bottom of each panel [3]. Image reproduced from Cordova et al. 2014. Magnetic force microscopy characterization of superparamagnetic iron oxide nanoparticles (SPIONs). Nano Biomed Eng 6(1): 1-12. Open Access.

Moreover, MFM allows the user to characterize magnetic nanoparticles in a non destructive manner without any special surface preparation or nanoparticle modification. In addition, MFM allows one to determine the magnetic moment of a single nanoparticle and how this measurement changes with nanoparticle size as well as probe distance from the sample, something which bulk magnetic analysis is not capable of. As all scanning probe methods it provides high resolution imaging of nanoparticle topography and quantitative analysis of nanoparticle size and density distribution, when nanoparticles are deposited on solid substrates. With successful development of MFM operating in liquid it opens excellent possibilities of studying magnetic nanoparticles in biologically relevant conditions, such as inside the cells, where AFM when AFM imaging is not working. Considering rapid development of novel applications of magnetic nanoparticles in medicine and biomedical nanotechnology, as therapeutic agents, contrast agents in MRI imaging and drug delivery [3] MFM characterization of nanoparticles becomes more valuable and desirable. Overall, MFM has proven itself to be an effective yet underused tool that offers great potential for the localization and characterization of magnetic nanoparticles.

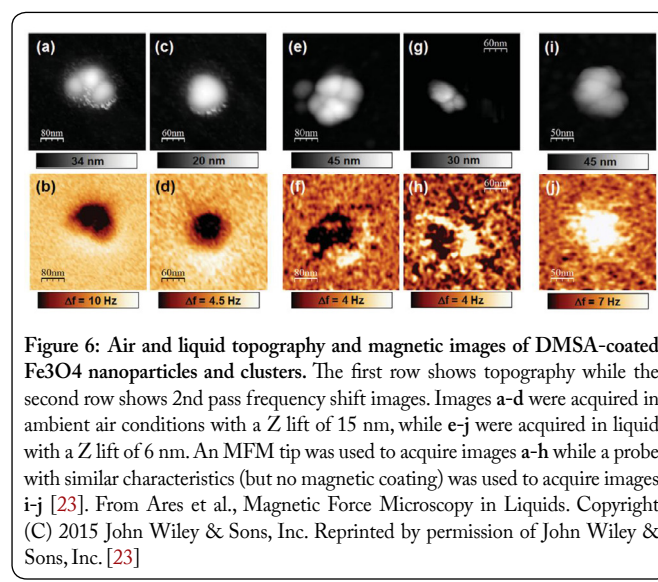

**Figure 6: Air and liquid topography and magnetic images of DMSA-coated Fe3O4 nanoparticles and clusters**. The first row shows topography while the second row shows 2nd pass frequency shift images. Images **a-d** were acquired in ambient air conditions with a Z lift of 15 nm, while **e-j** were acquired in liquid with a Z lift of 6 nm. An MFM tip was used to acquire images **a-h** while a probe with similar characteristics (but no magnetic coating) was used to acquire images **i-j** [23]. From Ares et al., Magnetic Force Microscopy in Liquids. Copyright (C) 2015 John Wiley & Sons, Inc. Reprinted by permission of John Wiley & Sons, Inc. [23]

## Conflict of Interest

The authors declare that they have no conflicts of interest with the contents of this article.

## Authors Contribution

Gustavo Cordova and Brenda Yasie Lee are contributed equally to this work.

## Acknowledgements

We greatly acknowledge the funding from Natural Science and Engineering Research Council of Canada (NSERC).